# Inverse probability of censoring weighting for visual predictive checks of time-to-event models with time-varying covariates

Christian Bartels[1] and Thomas Dumortier[1]

[1]Novartis Pharma AG, Basel, Switzerland

## 1 ABSTRACT

When constructing models to summarize clinical data to be used for simulations, it is good practice to evaluate the models for their capacity to reproduce the data. This can be done by means of Visual Predictive Checks (VPC), which consist of (1) several reproductions of the original study by simulation from the model under evaluation, (2) calculating estimates of interest for each simulated study and (3) comparing the distribution of those estimates with the estimate from the original study. This procedure is a generic method that is straightforward to apply, in general. Here we consider the application of the method to time to event data and consider the special case when a time-varying covariate is not known or cannot be approximated after event time. In this case, simulations cannot be conducted beyond the end of the follow-up time (event or censoring time) in the original study. Thus, the simulations must be censored at the end of the follow-up time. Since this censoring is not random, the standard KM estimates from the simulated studies and the resulting VPC will be biased. We propose to use inverse probability of censoring weighting (IPoC) method to correct the KM estimator for the simulated studies and obtain unbiased VPCs. For analyzing the Cantos study, the IPoC weighting as described here proved valuable and enabled the generation of VPCs to qualify PKPD models for simulations. Here, we use a generated data set, which allows illustration of the different situations and evaluation against the known truth.

## 2 INTRODUCTION

We have a study which follows patients until they experience an event, and we are interested in assessing whether the data could have been generated by a particular time-to-event model using a visual predictive check (VPC). The VPC consists of checking whether the Kaplan Meier (KM) estimate of the survival function calculated from the study data lies within its distribution under the model. The distribution is determined by Monte Carlo simulation, that is by repeatedly reproducing the original study by simulation from the model, and calculating the KM estimate for each of these simulated studies. If the KM estimate does not lie within its distribution under the model, it is considered unlikely that the model could have generated the study data.

Using VPCs to assess the appropriateness of models for simulations is a standard approach.[1-4] Here, we consider the special case when the model includes a time-varying covariate that is not known or cannot be predicted after the end of follow up in the original study. In this case, simulations cannot be conducted beyond this time point. An apparent solution is to censor the simulations at this time point. But as this additional censoring in the simulated studies is not random, the standard KM estimate from the simulated studies may be biased. In addition, this additional censoring in the simulated studies further reduces the number of patients that are at risk, which leads to an overestimation of the variability of the KM distribution.

To address those two problems, two new modifications of the established procedure to construct the visual predictive checks are proposed:

1. To correct for the bias, inverse probability of censoring (IPoC) weighting[5-7] has to be used for the simulated events.
2. To correct for the overestimation of the variability, simulations are carried out based on a marginal version of the model, which is again obtained by means of IPoC weighting.

# 3 METHODS

## 3.1 ORIGINAL STUDY AND MODELS

Each subject $i$ ($i = 1, ..., N$) from the original study is followed until it has experienced an event (at time $t_i$) or is censored (at time $c_i$), whichever comes first. The "end of follow up", denoted by $x_i^{obs}$, is thus equal to $min(t_i, c_i)$.

We have obtained or constructed by estimation a time-to-event regression model $M_0(Y)$ where $Y$ is a vector of covariates ($y_i$ for subject $i$) measured in the original study. We are not interested in how the model has been obtained or estimated. Instead, our goal is to evaluate $M_0(Y)$ using a VPC to assess how well it represents the event data from the original study.

Conceptually, the event times $t_i$ in the original study may be thought of having been generated by a heterogeneous time-to-event data generating process $M(Z)$, where the heterogeneity is induced by a prognostic factors $Z$. $M(Z)$ is characterized by a marginal survival function $S(t) = P(T > t)$. Usually, we do not know $S(t)$, or $M(Z)$.

We do also not know the process that generated the censoring times $c_i$ in the original study, denoted by $M_C$, but we assume here that it is homogenous (i.e., the same for all subjects) and independent from the time-to-event data generating process $M(Z)$. We have no direct interest in this process. Still, as we will see, it is necessary to incorporate the censoring process in the VPC. For this purpose, we have obtained or constructed by estimation from the observed censoring times in the original study a homogenous time-to-censoring model $M_{0C}$.

## 3.2 VPC ALGORITHM FOR TIME TO EVENT DATA

A VPC for time-to-event data is typically based on an estimator of the marginal survival function $S(t)$. The standard Kaplan-Meier (KM) non-parametric estimator, which is unbiased when event and censoring processes are independent, is the estimator usually selected for VPC. This estimator, that we denote by $\hat{S}(t)$, is defined as follows:

$$\hat{S}(t) = \begin{cases} 1 & if \ t < t_{[1]} \\ \prod_{\{r;\ t \geq t_{[r]}\}} \Delta S_r & if \ t \geq t_{[1]}, \end{cases}$$

where

$$\Delta S_r = 1 - \frac{d_r}{Y_r}$$

and $t_{[r]}$ is the r$^{th}$ event time observed in the original study, $\Delta S_r$ is the Kaplan Meier increment at time $t_{[r]}$, $Y_r$ is the number of patients in the risk set at time $t_{[r]}$, i.e., the patients who have neither experienced event nor been censored before that time, and $d_r$ is the number of patients with an event at time $t_{[r]}$.

The VPC assesses whether $\hat{S}(t)$, calculated from the original study data, lies within its distribution 'under the model $M_0(Y)$', i.e., its distribution if $M_0(Y)$ had generated the study data. If this is not the case, it is considered unlikely that the model $M_0(Y)$ has generated the study data. The distribution of the KM estimate under the model $M_0(Y)$ is obtained using a Monte Carlo approach. The procedure consists of repeated simulations of the study by random generation of the time-to-event data from the model $M_0(Y)$, to obtain a large number $J$ of 'simulated studies'. For each simulated study, $j$ ($j = 1, ..., J$), one event time, $t_{ij}$, is randomly generated for each subject, $i$ of the original study ($i = 1, ..., N$). The simulations are done based on the model $M_0(Y)$ to be evaluated and conditionally on the covariates, $y_i$, of each subject in the original study. The KM estimate is then calculated for each simulated study. The distribution of KM estimates across the simulated studies constitutes the Monte Carlo distribution of $\hat{S}(t)$ under the model $M_0(Y)$, and is used for the VPC assessment.

Censoring must also be taken into account when constructing the VPC. Censoring increases the variability of the KM estimate by reducing the risk set at later times (set of patients that are at risk of having an event at a given time). If this censoring is not taken into account in the simulated studies, then the Monte Carlo distribution of the KM estimate will under-represent its variability. In order to get the correct variability, the censoring process that occurred in the original study needs to be reproduced in the simulated studies by means of simulations from the censoring model $M_{0C}$ estimated from the original study. Specifically, censoring times are simulated from this model and compared to the simulated event-times to check whether the censoring or the event comes first and to decide based on this whether the patient gets censored or experiences an event in the simulated study. This procedure to obtain the KM estimates for each simulated study $j$, denoted by $\widehat{S_0}^j(t)$, is depicted in the 1$^{st}$ algorithm of Figure 1. With this procedure, the Monte Carlo distribution of the KM estimate should correctly represent its variability, and thus this distribution should be adequate for VPC assessment.

### 3.3 TIME-DEPENDENT COVARIATES MAY INTRODUCE BIAS IN VPCS

The above algorithm to produce VPCs to assess the appropriateness of models for simulations is a standard approach.[1-3,8] As mentioned in the introduction, we consider here the special case when the value of a time-varying covariate included in $M_0(Y)$ is not known or cannot be approximated after $x_i^{obs}$, the end of follow up (event or censoring time) in the original study. In this case, simulations of individual subjects cannot be conducted beyond $x_i^{obs}$, leading to additional censoring in the simulated studies when neither the simulated event time nor the simulated censoring time occur before $x_i^{obs}$. This additional censoring poses two problems.

First, the fact that subject can be censored in the simulated studies at time of event in the original study may introduce a dependency between censoring and event processes. This may result in the standard KM estimate of the survival function in the simulated studies to be biased. The bias is caused by the dependency between the two event processes $M(Z)$, which generated the original study events, and $M_0(Y)$, which generates the simulated studies' events, which occurs when those two processes share the same or at least stochastically dependent covariates ($Z$ and $Y$). Such a dependency is likely, since the model $M_0(Y)$ is supposed to capture important aspects of the original event process $M(Z)$. For example, subjects with higher risk of events, i.e., likely to experience events earlier in the original study and thus to be censored earlier in the simulated studies will be underrepresented in the risk set of the standard KM in the simulated studies. If the model captures correctly the higher risk of those subjects, then $Z$ and $Y$ are stochastically dependent and those subjects are also at high risk of event in the simulated studies. The underrepresentation of subjects with higher risk of events in the risk set will lead to an overestimation of $S(t)$ by the standard KM estimate in the simulated studies. Note that when the VPC is stratified, that bias in the standard KM estimate of the simulated studies is only present if the dependency condition is fulfilled within strata.

The second problem is that the additional source of censoring in the simulated studies (regardless of whether it is due to event or censoring in the original study) will further reduce the number of patients that are at risk, which leads to an increased variability compared to the KM estimate from the original study.

In summary, additional censoring in simulations at the end of follow up in the original study introduces biases into the Monte Carlo distribution of the standard KM estimate derived from the simulations, and makes this estimate inadequate for VPC assessments using the established standard approach.

### 3.4 CORRECTION OF POSSIBLE BIAS IN VPCS DUE TO TIME DEPENDENT COVARIATES

We propose to use inverse probability of censoring weighting (IPoC) to correct the KM estimate in the simulated studies with respect to the bias due to the additional censoring at $x_i^{obs}$, the end of the follow up in the original study.

IPoC has been established as an approach to correct for bias due to dependent censoring[5-7]. Dependent censoring occurs if the event process and the censoring process depend on some common covariates. If these covariates are all known, IPoC may be used to correct for the bias by conditioning on the covariates, which ensures conditional independence between the event process and the censoring process, as follows[5]: At each event time, the probability of not being censored by that time is calculated dependent on the covariates for each subject who remains in the study. The increment in the KM estimate at this time is then modified by weighting each subject by the inverse of this probability. As such, groups of subjects that are underrepresented at that time get a higher weight and those subjects of the group that remain in the study can represent all the subjects that would have remained in the study without the dependent censoring.

In our case, the events in the simulated studies are generated from the model $M_0(Y)$, and censoring is a composite process made of the earliest of 1) an event in the original study generated from $M(Z)$, (2) a censoring event in the original study generated from $M_C$, and (3) a simulated censoring event generated from $M_{0C}$. To apply IPoC in each simulated study, it is necessary to calculate, for all subjects $i$ ($i = 1..N$) who have survived until time t in the simulated study, the probability of not having been censored in that simulated study. Conditioning this probability on all covariates Y of the model $M_0(Y)$ ensures that the events in a simulated study are conditionally independent to the composite censoring process. This conditional independence ensures the validity of IPoC weighting as discussed in the previous paragraph.

The conditional probability of not having being censored in a simulated study can be calculated as, by assumption, the event process and the censoring process of the original study, $M(Z)$ and $M_C$, are independent, and, by construction, these two processes $M(Z)$ and $M_C$ are also independent from the censoring model $M_{0C}$. Consequently, under the event model $M_0(Y)$ and the assumption that the censoring model $M_{0C}$ is a good approximation of the true censoring process $M_C$, this conditional probability can be approximated at time $t$ by $S_0(t, y_i)S_{0C}(t)S_{0C}(t)$, where $S_0(t, y_i)$ is the survival function characterizing the event model $M_0(Y)$ evaluated at time $t$ for subject $i$ with covariate $Y = y_i$ and $S_{0C}(t)$ is the survival function characterizing the censoring model $M_{0C}$ evaluated at time $t$. The IPoC weights are therefore equal to $W_i(t) = \left(S_0(t, y_i)S_{0C}(t)S_{0C}(t)\right)^{-1}$.

The KM is modified to include the weights as follows:

$$\widehat{S_0}^{j,mod}(t) = \begin{cases} 1 & if \ t < t_{[1]j} \\ \prod_{\{r; t \geq t_{[r]j}\}} \Delta S_r^j & if \ t \geq t_{[1]j} \end{cases},$$

where

$$\Delta S_r^j = 1 - \frac{\sum_{\{i; \delta_{ij}(t_{[r]j})=1\}} W_i(t_{[r]j})}{\sum_{k \in R_r^j} W_k(t_{[r]j})}$$

and $t_{[r]j}$ is the r[th] event time 'observed' in simulated study $j$, $\Delta S_r^j$ is the increment of this modified Kaplan Meier estimator at time $t_{[r]j}$, $R_r^j$ is the risk set at time $t_{[r]j}$ in the simulated study $j$, i.e., the set of subjects who have not experienced event nor been censored before that time in that simulated study, and $\delta_{ij}(t)$ is equal to 1 if subject $i$ has 'experienced' event in the simulated study $j$ at time $t$ and to 0 otherwise. Note that due to the homogeneity of the censoring process, a simplified weight equal to $W_i(t) = S_0(t, y_i)^{-1}$ gives the same results.

This VPC procedure, using $\widehat{S_0}^{j,mod}(t)$ instead of $\widehat{S_0}^j(t)$ to estimate the marginal survival function in the simulated studies, is schematically depicted in Figure 1, 2[nd] algorithm.

## 3.5 CORRECTING THE VARIABILITY OF THE VPC

Even though $\widehat{S_0}^{j,mod}(t)$ is an unbiased estimate of the marginal survival function under $M_0(Y)$, its Monte Carlo distribution has a higher variability than the variability that would be obtained by repeated execution of the trial, due to the additional censoring at $x_i^{obs}$. To address this second problem, we propose to generate simulated studies from $M_0$, the marginal version of $M_0(Y)$, that is defined by its survival function $S_0(t) = \sum_{i=1:N} S_0(t, y_i)/N$. Therefore, instead of simulating one event for each of the N patients from the model $M_0(Y)$, i.e., as a function of their covariate $y_i$, we simulate events of the N patients from the marginal model. As in the previous section, $S_0(t, y_i)$ and thus $S_0(t)$ cannot be calculated after patient $i$ is censored or experience event in the original study. However $S_0(t)$ can be approximated by averaging the IPoC corrected estimates $\widehat{S}_0^{j,mod}(t)$ for a large number of simulated studies. We denote this average by $\overline{\widehat{S}_0}^{mod}(t)$:

$$\overline{\widehat{S}_0}^{mod}(t) = \sum_{j=1:J} \widehat{S}_0^{j,mod}(t)/J,$$

and derive the corresponding model that we denote by $\widehat{M}_0$.

The simulated studies that are used to calculate the MC distribution of the KM estimate are generated as with the 1st algorithm, except that all times of events are simulated from $\widehat{M}_0$. The procedure is depicted in Figure 1, 3rd algorithm. Since an unbiased marginal model is used for the simulations, and since the simulated event times are independent from the simulated censoring times, the standard KM estimates for the simulations are unbiased. Further, since the simulation can be conducted beyond the end of follow up time in the original study, the Monte Carlo variability is expectedly adequate.

The different VPC methods are illustrated by an example in the next sections.

**Figure 1    Calculation of the survival function estimate for the simulated studies**

For,

- $x_i^{obs}$: Event or censoring time observed in the original study for subject $i$ ($i=1..N$)
- $M_{C0}$: Assumed censoring model
- $M_0(Y)$: Assumed event model conditional on the covariate $Y$
- $S_0(t, y_i)$: Survival function corresponding to model $M_0(Y)$ evaluated at time $t$ for subject $i$ with covariate $Y = y_i$.
- $c_{i,j}$ and $t_{i,j}$: Simulated censoring and event times for subject $i$ in simulated study $j$ ($j = 1..J$)
- $(x_{i,j}, \delta_{i,j})$: $\delta_{i,j}$ is a flag indicating whether subject $i$ in simulated study $j$ is censored ($\delta_{i,j} = 0$) or experiences an event ($\delta_{i,j} = 1$), and $x_{i,j}$ is the corresponding time
- $\widehat{M}_0$ and $\widehat{S}_0^{mod}(t)$: marginal model and corresponding survival function

1st algorithm – Standard VPC procedure:

For each simulated study $j = 1..J$

- For each subject $i = 1..N$
  - Generate censoring time $c_{i,j} \sim M_{C0}$
  - Generate event time $t_{i,j} \sim M_0(Y)$ for $Y = y_i$
  - Derive $(x_{i,j}, \delta_{i,j}) =$
    $(c_{i,j}, 0), \quad \text{if } c_{i,j} < t_{i,j}$
    $(t_{i,j}, 1), \quad \text{if } c_{i,s} \geq t_{i,s}$
- Calculate[1)] the standard KM estimate $\widehat{S}_0^j(t)$ from $(x_{i,j}, \delta_{i,j})_{i=1..N}$

2nd algorithm – IPoC weighting:

For each simulated study $j = 1..J$

- For each patient $i = 1..N$
  - Generate censoring times $c_{i,j} \sim M_{C0}$
  - Generate event times $t_{i,j} \sim M_0(Y)$ for $Y = y_i$
  - Derive $(x_{i,j}, \delta_{i,j})$
    $(x_i^{obs}, 0), \quad \text{if } c_{i,j} > x_i^{obs} \text{ and } t_{i,j} > x_i^{obs}$
    $(c_{i,j}, 0), \quad \text{if } c_{i,j} < t_{i,j} \text{ and } c_{i,j} \leq x_i^{obs}$
    $(t_{i,j}, 1), \quad \text{if } c_{i,j} \geq t_{i,j} \text{ and } t_{i,j} \leq x_j^{obs}$
- Calculate[1)] IPoC KM estimate $\widehat{S}_0^{j,mod}(t)$ from $(x_{i,j}, \delta_{i,j}, S_0(.,y_i))_{i=1..N}$ that corrects for the additional censoring[1)]

3rd algorithm – IpoC weighting plus correction of variability:

Prior to simulations, approximate $S_0(t)$ by $\overline{\widehat{S}_0}^{mod}(t)$ which is equal to the mean of $\widehat{S}_0^{j,mod}(t)$, determined with the 2nd algorithm, across the $J$ simulated studies; derive the corresponding model $\widehat{M}_0$.

For each simulated study $j = 1..J$

- For each patient $i = 1..N$
    - Generate censoring times $c_{i,j} \sim M_{C0}$
    - Generate event times $t_{i,j} \sim \widehat{M}_0$
    - Derive $(x_{i,j}, \delta_{i,j}) =$
      $(c_{i,j}, 0),\quad if\ c_{i,j} < t_{i,j}$
      $(t_{i,j}, 1),\quad if\ c_{i,j} \geq t_{i,j}$
- Calculate[1] the standard KM estimate $\widehat{S}_0^j(t)$ from $(x_{i,j}, \delta_{i,j})_{i=1..N}$

Footnote to Figure 1:

[1] The equations to calculate the KM estimate proposed in Sections 3.2 and 3.4 handle ties, i.e, events that occur at the same time. To handle ties, the times series $t_{[r]j}$ at which events occur needs to be determined from the simulated events $(x_{i,j}, \delta_{i,j})_{i=1..N}$, together with the indicator $\delta_{ij}(t_{[r]j})$ whether a patient $i$ has an event at time $t_{[r]j}$ and the corresponding risk set, $R_r^j$ of subjects who have not experienced an event nor been censored before that time in the simulated study.

## 3.6 EXAMPLE FOR ILLUSTRATION

The work described here was motivated and has been used for the development of a PKPD model for some of the data of the Cantos study.[9] For this study, it was not clear how to simulate dosing at times after which the patients discontinued the study. For the Cantos data, the IPoC weighting as described here proved valuable and enabled generation of VPCs to qualify the PKPD models for simulations.

Here an artificial data set is used for illustration. The data and the model are such that simulations beyond the time of the event or censoring are possible. Having such data, different situations can be illustrated

1) Standard VPC for a model that supports simulations beyond the time of the observed event/censoring (Figure 3)
2) Biased result, if using the standard VPC for a model that does not support simulations beyond the time of the observed event/censoring (Figure 4)
3) Corrected, unbiased result, if using IPoC VPC for a model that does not support simulations beyond the time of the observed event/censoring
    a. With variability as obtained from repeated simulations analyzed with IPoC (Figure 5)
    b. Variability as obtained by repeated simulations from the marginal survival model (Figure 6)

To simulate the data to be analyzed, the data generating process $M(Z)$ of the original study, which is usually not know, was assumed to be as follows:

- The original study included 1000 high risk and 1000 low risk patients.
- In each risk group, 500 patients were assigned to placebo and 500 patients to active treatment.
- The event times of the 2000 patients were assumed to follow an exponential distribution with hazard function $h(TRT) = \lambda_{TRT,g}$ for risk group $g$ and treatment $TRT$ with the hazard rates $\lambda_{TRT,g}$ listed in Table 1; the treatment is assumed to decrease the hazard in low risk patients and have no effect in high risk patients.
- Censoring was assumed to be independent from the event process and to follow a Weibull distribution with scale and shape parameters equal to 2 and 5, respectively.

The event and censoring times of the original study were simulated from these distributions; the simulation is summarized via a KM estimate displayed in Fig. 2 and is assumed to be the actual data from the original study in the evaluation of the VPC that follows.

Table 1. Hazard $\lambda_{tg}$ used to simulate the event times of the 2000 patients included in the original study

| Hazard | Low risk patients | High risk patients |
| --- | --- | --- |
| **Active** | 0.05 | 2.0 |
| **Placebo** | 0.2 | 2.0 |

Figure 2      KM estimates from the original study

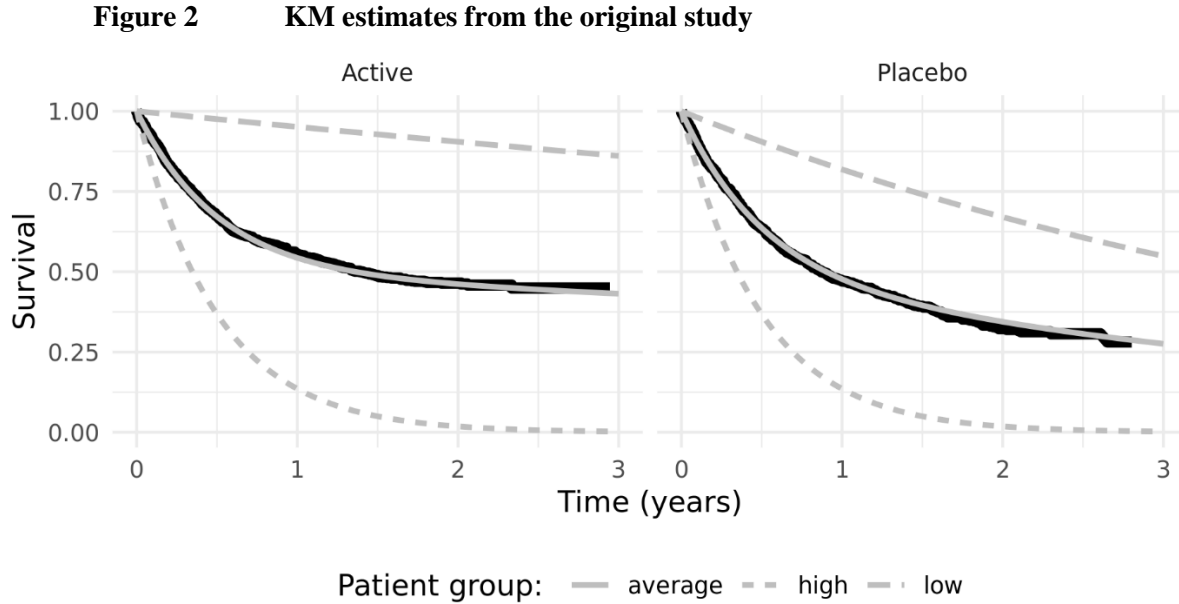

*Footnote: Black lines: KM estimated from the original study data which was simulated from the event process $M(Z)$ with rate displayed in Table 1 and the censoring described above. Grey lines illustrate the data generating process for low and high risk patients (dashed lines) and average patients (continuous line).*

The model $M_0(Y)$ subject to our evaluation is a Cox proportional hazard model[10] with the same explanatory covariates as $M(Z)$ (treatment, patient group and their interaction) and was estimated from the data of the (simulated) original study. Although this Cox model differs from the (true) data generating process $M(Z)$ used to generate the data (Table 1), it is a more general model than the data generating process; therefore, a valid VPC procedure should not detect model misspecifications and just illustrate uncertainty due to random variability. The model used active high-risk patients as reference. The model captured the simulated clinical data adequately and estimated the following effects

- Low-risk patients had 0.025 times lower hazard than high-risk patients,
- High-risk patients on placebo were estimated to have a 1.09 times higher hazard than those on active treatment,
- Low-risk patients on placebo were estimated to have a 3.9 times higher hazard than those on active treatment

The censoring model, $M_{C0}$, is a Cox proportional hazard model without any explanatory variable and was as the survival model estimated from the data of the original study.

All the analyses have been implemented in R[11] version 3.6.1 using the survival package[12] version 3.1-7.

# 4 RESULTS

When being able to simulate event and censoring times for all patients, even after their end of follow-up in the original study, the Standard VPC procedure (1st algorithm in Figure 1) works. Figure 3 shows the KM estimates of the original study (dashed black line) together with the VPC represented by the mean and 90% prediction interval obtained by Monte Carlo simulation (grey band with white line). Since the model under evaluation is consistent with the data generating process, the predictions based on the model match the study data. In each treatment group, the KM estimate of the original study data (dashed black line) lies within the prediction interval constructed under the model. This supports that the model could have generated the study data.

**Figure 3**     Reference of visual predictive checks for the two treatment groups

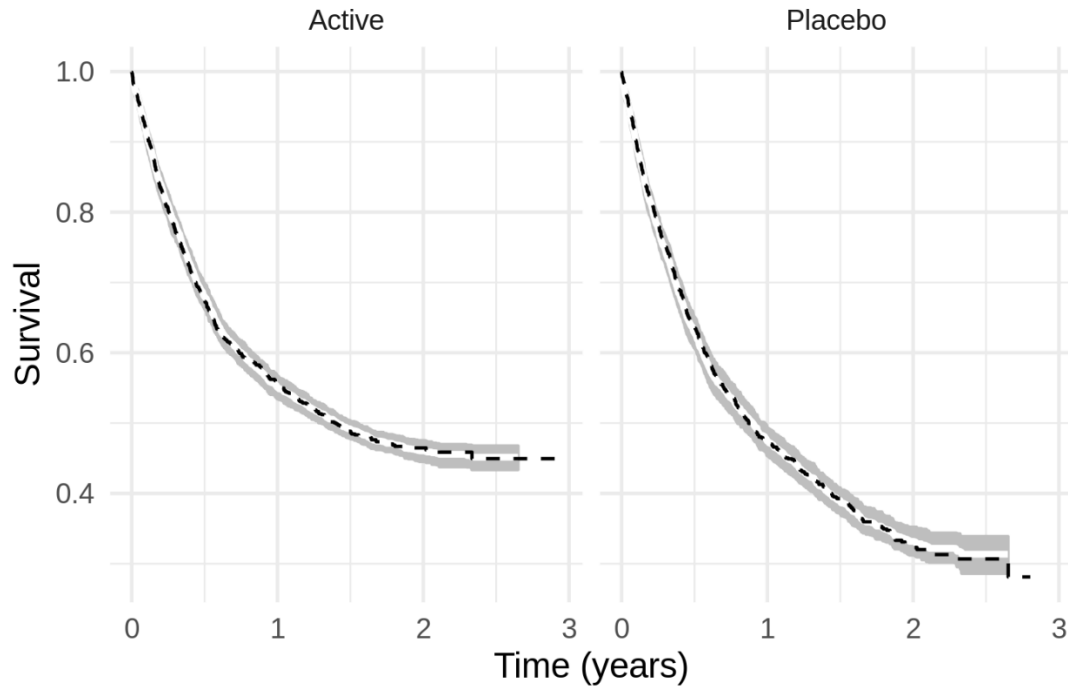

*Footnote: Dashed line is the KM estimate $\hat{S}(t)$ derived from the original study data. The white continuous line with the shade area are the VPC with the mean and 5% to 95% percentile of the distribution of standard KM estimates $\widehat{S_0}^j(t)$ across simulated studies j (j=1..J) under the model, $M_0(Y)$ of interest.*

However, when the simulations must be censored at the time of the end of the follow up (time of event or censoring) in the original study, the standard KM estimates of the simulated trials do not match the KM of the original study data (Figure 4). The difference is pronounced for the patients on active treatment for which differences in hazard of the two risk groups are large.

As laid out, those standard KM estimates of the simulated studies are biased, due to dependent censoring in the simulations at time of event in the original study. Because of the dependent censoring, patients at higher risk of events (i.e., patients in the high risk group) are under-represented in the simulations at later times. This unequal representation leads to a bias when using the default VPC procedure that assumes that events are drawn for all patients at all times t according to their predicted survival $S_0(t, y_i)$.

Note again that independent censoring was assumed for the original study results. As such the estimates of the study results (black dashed lines) are not biased. The bias is only present in the estimates from the simulations for the VPC and is due to additional censoring in the simulated studies at end of follow up time in the original study, and the bias in these estimates is visible as difference between the estimates from the study and the VPC.

**Figure 4: Visual predictive checks with censoring of simulations at the observed end of follow up - VPC stratified by treatment**

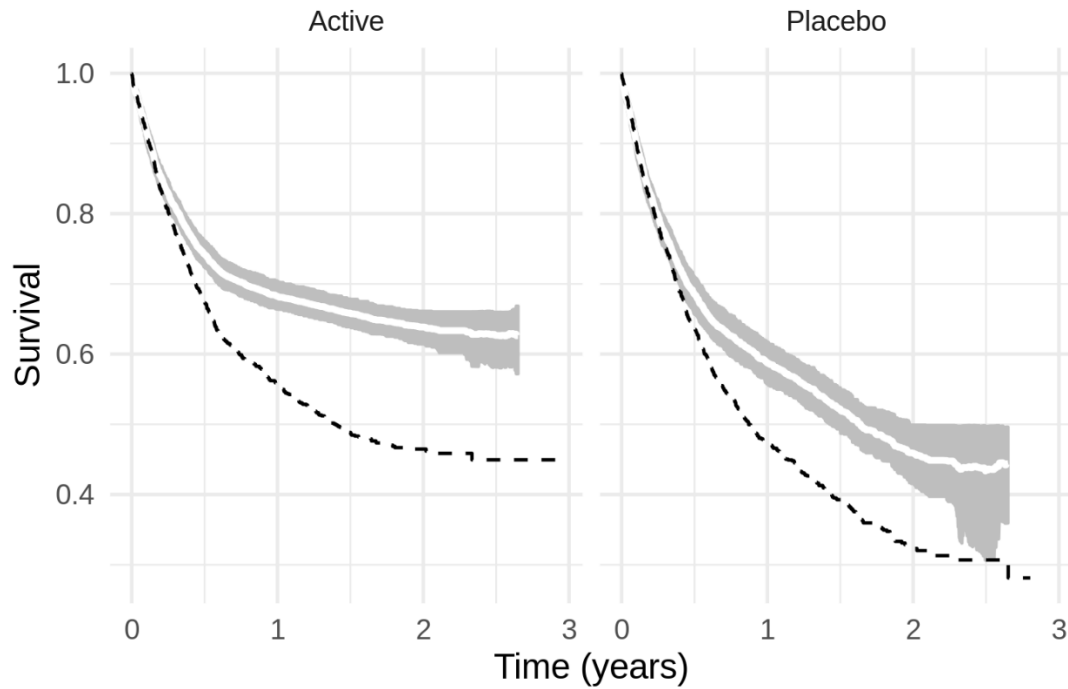

*Footnote: Dashed line is the KM estimate $\hat{S}(t)$ derived from the original study data. The white continuous line with the shade area are the mean and 5% to 95% percentile of the distribution of standard KM estimates under the model $M_0(Y)$ of interest.*

IPoC weighs the contribution of each subject by the inverse of its probability of not being censored in the simulations. As such, groups of subjects that are underrepresented at time *t* get a higher weight and those subjects of the group that remain in the study can represent all the subjects that would have remained in the simulation without the dependent censoring.

Fig. 5 shows that applying IPoC to the simulated trials used for Figure 4 results in unbiased modified KM estimates (2nd algorithm in Figure 1). Indeed, the mean of the modified KM estimates matches again the study data.

**Figure 5: Visual predictive checks using IPoC to correct KM estimates of the simulated trials**

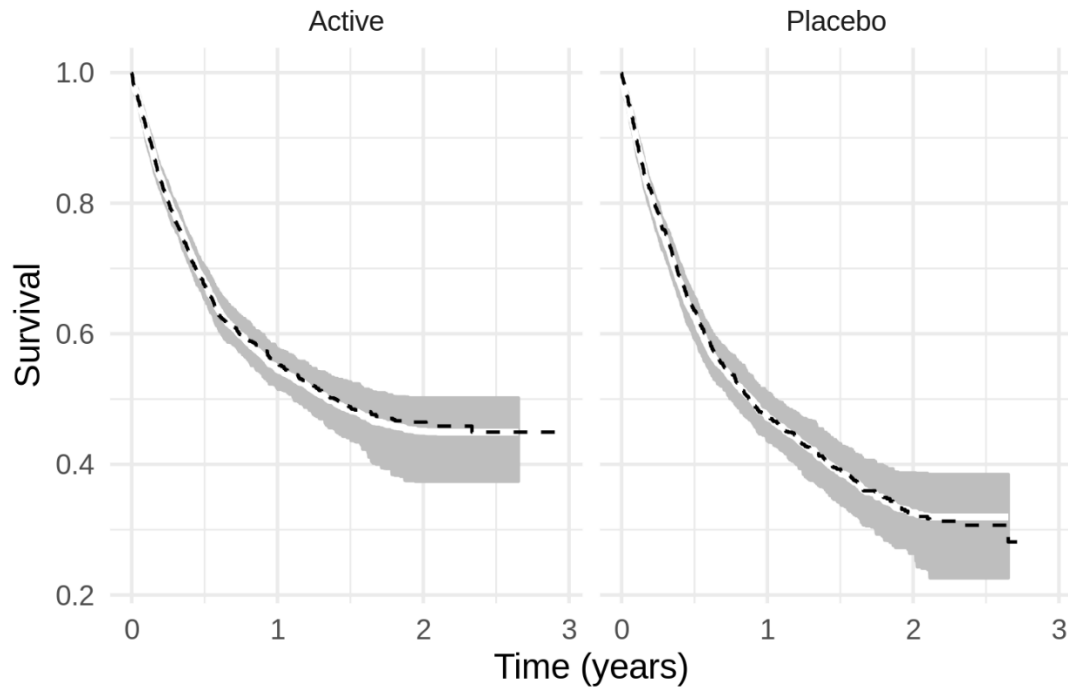

*Footnote: Dashed line is the KM estimate $\hat{S}(t)$ derived from the study data. The white continuous line with the shade area are the mean and 5% to 95% percentile of the distribution of the modified KM estimates $\widehat{S_0}^{j,mod}(t)$ across simulated studies j (j=1..J) under the model, $M_0(Y)$ of interest.*

However, compared to the situation in which events could be generated for all subjects after their end of the follow up in the original study (Figure 3), the estimate of the variability of the simulated trials in Figure 5 is too large. Consequently, the VPC procedure has low power to reject an incorrect model under evaluation.

The solution is to simulate events from the marginal model that we approximate using IPoC and that we denoted above by $\overline{\overline{S_0}}^{mod}(t)$. Events can be simulated from this marginal model up to the end of the study. The simulated events are compared as in the default approach with the simulated censoring events (3rd algorithm in Figure 1).

Unbiasedness and adequacy of the variability is confirmed in Figure 6. In particular the Monte Carlo variability is similar to that obtained in Figure 3.

**Figure 6: Visual predictive check simulating from IPoC corrected estimate of mean survival**

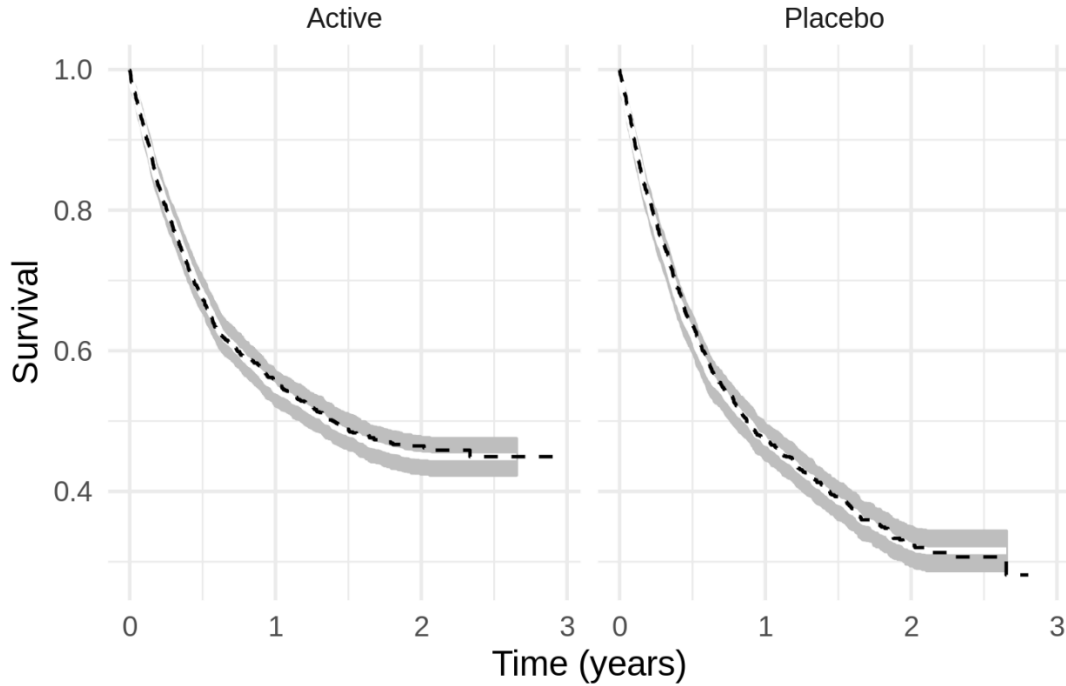

*Footnote: Dashed line is the KM estimate $\hat{S}(t)$ derived from the study data. The white continuous line with the shade area are the mean and 5% to 95% percentile of the distribution of distribution of standard KM estimates $\widehat{S_0}^j(t)$ across simulated studies j (j=1..J) under the model, $M_0(Y)$ of interest.*

# 5 DISCUSSION

IPoC weighting is an established method for the analysis of time to event data. [5-7] IPoC weighting is used to correct for biases that may occur with statistical analyses of study results, when censoring and events observed in the study are not independent. Here, IPoC weighting is used differently to correct simulation results required for the construction of the VPCs. Indeed, even if the censoring and events observed in the original study are independent, this correction may be required in the simulated studies if the model to be evaluated does not support simulations beyond the time of the observed event/censoring in the original study.

Similarly, VPCs are an established approach to qualify models of time to event data for simulations.[1-3,8] The bias that occurs with VPCs when a time-varying covariate is not known or cannot be approximated after the end of follow up (event or censoring time) has not received attention so far. Here, we illustrated that this additional censoring leads (i) to a bias in the standard KM estimate from the simulated study and (ii) further reduces the number of patients that are at risk, which leads to an overestimation of the variability. To address these two problems, we proposed two new modifications of the established procedure to construct the VPC. Ignoring these biases or not having tools available to correct for them may have important consequences. Ignoring the biases may lead to qualification of a model for simulations that does not represent the clinical data, or to the rejection of models that represent the clinical data. Noticing the presence of the bias but not having the tools to correct for them would make model qualification more difficult or impossible.

Our approach of using IPoC weighing to modify the KM estimate of the simulated studies and approximating the marginal survival function of the model under evaluation is simple to implement in that no additional model is required other than the survival model that should be assessed using the VPC. An alternative is to establish a model for the time-varying covariates, in which case this model can be used to predict the subjects' covariate time-profiles in the simulated study after end of follow up in the original study and the standard VPC procedure is valid. But it should be noted that modeling the covariate process on its own may yield biased inference in case this process is internal[13] (or when event and covariate processes share a random effect) as censoring at time of event cause non-ignorable missingness for the covariate process[14]. Instead it may be necessary to jointly model the covariate and the event process, which is technically demanding[14]. This complexity may not matter when the covariate process is interesting in itself and a joint model is used to understand the causal relation between those two processes. However, if one does not want to face those difficulties for the sole purpose of constructing unbiased VPCs, the approaches proposed here should be sufficient. Finally, it should be noted that the covariate process may depend on factors difficult to predict, for instance in a Therapeutic Drug Monitoring setting when it is difficult to predict physicians' decision to adjust the dose and thus the drug concentration time-course. In this case also, our approaches are at least convenient.

# 6 CONCLUSION

VPCs are a standard tool to assess the appropriateness of a model for simulations that is also routinely used for time to first event data. If the model to be evaluated cannot be used to simulate events beyond the end of the follow up from the original study, e.g., due to time-varying covariates, informative censoring occurs, and standard KM estimates based on simulated events may be biased. We showed that using IPoC, it is possible to correct for this bias, and to recover valid VPCs to enable the assessment of the model for simulation purposes.